# Adaptive Blind Image Watermarking Using Fuzzy Inference System Based on Human Visual Perception

Maedeh Jamali[1], Shima Rafiei[1], S.M. Reza Soroushmehr[2], Nader Karimi[1], Shahram Shirani[3], Kayvan Najarian[2], Shadrokh Samavi[1,2,3]

[1]Department of Electrical and Computer Engineering, Isfahan University of Technology, Isfahan 84156-83111, Iran.
[2]Department of Computational Medicine and Bioinformatics, University of Michigan, Ann Arbor, 48109 U.S.A.
[3]Department of Electrical and Computer Engineering, McMaster University, Hamilton, L8S 4L8, Canada.

*Abstract*— Development of digital content has increased the necessity of copyright protection by means of watermarking. Imperceptibility and robustness are two important features of watermarking algorithms. The goal of watermarking methods is to satisfy the tradeoff between these two contradicting characteristics. Recently watermarking methods in transform domains have displayed favorable results. In this paper, we present an adaptive blind watermarking method which has high transparency in areas that are important to human visual system. We propose a fuzzy system for adaptive control of the embedding strength factor. Features such as saliency, intensity, and edge-concentration, are used as fuzzy attributes. Redundant embedding in discrete cosine transform (DCT) of wavelet domain has increased the robustness of our method. Experimental results show the efficiency of the proposed method and better results are obtained as compared to comparable methods with same size of watermark logo.

## I. Introduction

Digital imaging has developed quickly in the last two decades. Because of easy access to internet, sharing digital media has become easier and faster in recent years. Unlike their analog counterparts, digital contents can be copied with the same quality as the original one. As a result of this extensive development, access to image editing applications can threaten the integrity of digital media. Increasing the possibility of digital tampering has emphasized the need for sophisticated techniques of copyright protection and prevention of unauthorized copying and distribution [1].

One usual approach is the hashing of the original image. If the hash output in receiver side is the same as the one transmitted from original image, then the image is considered as unchanged [2-3]. However, these methods require secure channel for each image transmission. Due to rare availability of such requirements, watermarking is the prevalent means of copyright protection. Watermarking is one of the most common and trusty ways to solve this problem [4]. In this method, the owner's copyright information is embedded into an image that later can be extracted from the watermarked image. Potdar *et al.* reviewed some watermarking methods. They referred to robustness, imperceptibility and capacity as three essential factors that should be considered in digital watermarking. The efficiency of a watermarking technique relies highly on the trade-off among these features. The watermarking methods can be classified in different ways based on the embedding domain, embedding method and the extraction process. Watermark methods are classified into three classes based on whether they need to have the original image in the extraction phase or not. These categories include blind, semi-blind and non-blind watermarking methods. Blind watermarking systems are more attractive than the other two techniques. This is due to the fact that blind schemes do not need the original image in their extraction phase.

Another type of classification for watermark methods is based on the embedding domain. Watermarking can be categorized into two groups of spatial and transform domain techniques. In spatial domain methods, information is directly embedded in image pixels. However, these methods usually are not necessarily robust against usual image processing attacks. On the other hand, frequency domain algorithms are more robust in most cases [5]. Some spatial domain methods are proposed in [6-8]. Lin *et al.* [6] proposed watermarking scheme using the $1/T$ rate *forward error correction* (FEC) where $T$ is the data redundancy rate. For better security, the watermark logo is mixed with noise bits. Also, they XOR the watermark with a binary feature value of the image by 1/T rate FEC. Their algorithm has a blind extraction phase and the watermark bits are determined by majority voting. In [7], the embedding phase starts by preprocessing of the cover image by a Gaussian low-pass filter and using a secret key, a number of gray levels are randomly selected. A histogram of filtered image based on these selected gray levels is made. The novelty of their work is using histogram-shape-related index to select pixel groups with highest number of pixels. Also, a safe band is constructed between chosen and non-chosen pixel groups. The watermark is then inserted into the chosen pixel groups. However, the capacity of this method is low. Work of [8] is another spatial domain technique that a set of affine invariants is derived from Legendre moments. These affine invariants are used in embedding.

Transform methods embed information into frequency coefficients of the original image. Hence, they can be more powerful than the spatial domain groups and better preserve image imperceptibility [5]. Many different transform techniques, such as discrete Furrier transform (DFT) [9], *discrete cosine transform* (DCT) [10-18], *discrete wavelet transform* (DWT) [19-32] and *Contourlet transform* (CT) [33-40] have been used for digital image watermarking. In [9], the watermark is added in middle frequencies of DFT and it has circular symmetry. Suhail and Obaidat [10] propose a method in which an input image is divided into different parts based on Voronoi algorithm. After that the DCT of these segmented parts hide a sequence of real numbers. Authors of [11] apply DCT to image blocks and a low resolution approximate image is formed using the DC coefficient of each block. Eventually, embedding is done by adding watermark to high frequencies of the reconstructed image. Bors and Pitas [12] propose a DCT based method where the input image is divided into $8 \times 8$ blocks. Then block is transferred into DCT and some blocks based on a Gaussian network classifier are selected. Finally, DCT coefficients are modified based on the watermark data. Bors and Pitas [13] propose a watermarking method and show that different attacks tend to change different parts of frequency spectrum. Hence, by detecting how different attacks affect the image, more precise extraction is performed. They propose an attack classification method to recognize regions of the frequency domain that are less damaged. Hence, the watermark can be extracted from regions with less damage. Huang and Guan proposed a hybrid DCT and *singular value decomposition* (SVD) based watermarking [14]. In their method, SVD and DCT are applied on the watermark and original images respectively and the singular values of the watermark are embedded into DCT coefficients of the original image. Another DCT-SVD based algorithm is proposed in [15]. First a mask of the original image is built using luminance mask. The embedding process is done by modifying singular values of DCT of the original image with singular values of a produced mask. The control parameter is found by genetic algorithm. In [16], an adaptive watermarking algorithm for medical images is proposed which uses *fuzzy inference system* (FIS) and characteristics of *human visual system* (HVS). The embedding is done in the wavelet domain. The HVS model contains luminance, texture, and frequency sensitivities. The algorithm of [17] uses neuro-fuzzy method consisting of back propagation neural networks and fuzzy logic techniques for embedding and extraction procedures. In their algorithm HVS parameters are fed to the system and the output of fuzzy-neural is used as the strength factor for the watermark embedding. In [18] three fuzzy inference models are used for formation of the watermark strength factor, which uses inputs based on HVS.

A survey of DWT base watermarking is explained in [19]. DWT is widely used in watermarking algorithms. The middle or high frequency regions of the coefficients are usually used for embedding [20]. Xia *et al.* [21] introduce a watermarking method based on DWT. They add pseudo-random codes to large coefficients at high and middle frequency bands, but the high frequency band is not robust against attacks such as JPEG. In [22] a DWT-SVD method is introduced. They embed watermark in singular values of the wavelet transform's sub-bands of the original image. Authors of [23] propose wavelet tree clustering for data hiding. Wavelet trees are produced by distance vector. These trees are classified into two clusters: one denotes a watermark bit of 1 and the other shows 0. Statistical difference and the distance vector of a wavelet tree are compared for the extraction of the embedded bit. In [24] a blind watermarking method is proposed which uses quantization of maximum wavelet coefficient. Wavelet coefficients of the input image are categorized into different blocks. The embedding is done in different sub-bands. They add various energies to maximum coefficients so that, this always remain maximum in each block. This method has the drawback of having low normalized correlation (NC) against intense JPEG attacks. In [25] and [26] the combination of DCT and DWT domains is used. The watermark is embedded in the middle frequency coefficients in DCT domain of three DWT levels of the LL band of the original image. In some papers combination of DWT and SVD are introduced as watermarking methods. Authors of [27] apply SVD to all frequency bands in DWT of the original image for the watermarking purpose. In [28] a quantization based watermarking is proposed. They embed a watermark bit by quantizing the angles of considerable gradient vectors in different wavelet scales. Authors of [29] have proposed geometrical model for embedding to generate a tradeoff between robustness and transparency. They use eight samples of wavelet approximation coefficients from each image block and built two-line segments in a two-dimensional space. The proposed method of [30] is based on context modeling and fuzzy inference filter which are to determine coefficients with large entropy in coarser DWT sub-bands for watermark embedding. The algorithm of [31] utilizes Fuzzy logic to obtain a perceptual weighting factor for each wavelet coefficient for embedding at different scales of an image.

Some researchers have used different types of transforms such as CT, DCT, and Hadamard transform for embedding process. In [33] proposed to embed watermark into different bands of CT to increase robustness. Authors of [34] introduce an adaptive blind-based watermarking in which the watermark is embedded in DCT coefficients of CT. They apply two-level CT to the cover image. In the first level, the approximate image is divided into blocks. Important edges of each block are determined using their proposed edge detection method. Parts of an image with high concentration of edges are considered as candidate parts for strong embedding. Some portions of the second level are also concatenating with mentioned blocks. The entropy of blocks and some other criterions of each block determine an adaptive strength factor for that block. Then DCT transform of blocks is used for embedding. Authors of [35] use Hadamard transform for obtaining robust and low complexity embedding mechanism. In [36] a hybrid of CT and DCT is used. The strength factor severity is calculated based on CT complexity of each block. In [37] CT domain is used for embedding. They embed the watermark in DCT coefficients of CT blocks for more robustness. A major drawback of this method is its poor performance against attacks such as salt & pepper noise, Gaussian noise and JPEG compression. The method of [38] uses maximum likelihood method based on normal inverse Gaussian (NIG) distribution for watermark extraction. Authors of [39] have proposed a watermark method based on a sample projection technique. They use low frequency components of image blocks for embedding in order to achieve more robustness against different attacks. They use four samples of approximation coefficients of image blocks to build line-segment in 2D space. The slop of this line segment is used for embedding. In [40] a contourlet-based watermarking is proposed. They use nine samples of approximation coefficients of an image block to build a plane in 3D space. Embedding is done by changing the dihedral angle between created plane and the x-y plane.

Fuzzy logic models have been used in watermarking algorithms to obtain appropriate weighting factors to embed watermarks. Most FIS and HVS-based algorithms in the literature heuristically find membership functions as their input features. The membership functions are not tuned subjectively by users or objectively by the real distributions of the features. Furthermore, none of the mentioned algorithms consider saliency, based on human fixation locations, which could be an effective FIS input.

In this paper, we propose a new adaptive blind watermarking method in DWT domain. This method uses a hybrid of DWT and DCT. We first apply one level 2D DWT on the original image. A second 2D DWT is applied on the vertical, horizontal and diagonal sub-bands from first level DWT. In order to increase robustness against attacks, the watermark is embedded in the DCT coefficients of the wavelet blocks. To achieve a tradeoff between imperceptibility and robustness, we use different strength factors to embed in different blocks. These strength factors are computed by a fuzzy system. We use fuzzy system because it models the uncertainty and inference similar to a human. Saliency, edge-concentration and intensity are the three fuzzy inputs that form the fuzzy term set. Each term set has its special membership function. After calculating the strength factor base on the fuzzy system, the embedding is performed. In brief, we propose a novel method using a fuzzy system for finding regions where HVS is not sensitive to their changes. Also for increasing robustness, we redundantly embed the watermark and a voting is performed in the extraction stage. Embedding is done with different strengths in different wavelet sub-bands.

The rest of this paper is organized as follows. Section II contains the details of the proposed method. In section III the experimental results are presented. Finally, the paper is concluded in section IV.

## II. PROPOSED METHOD

In this section, we explain our proposed watermarking method. We first explain our feature extraction phase and how the fuzzy system inputs are prepared. Figure 1 shows the main steps of our fuzzy strength factor map generation. After that each fuzzy model, the embedding and extraction phases are explained.

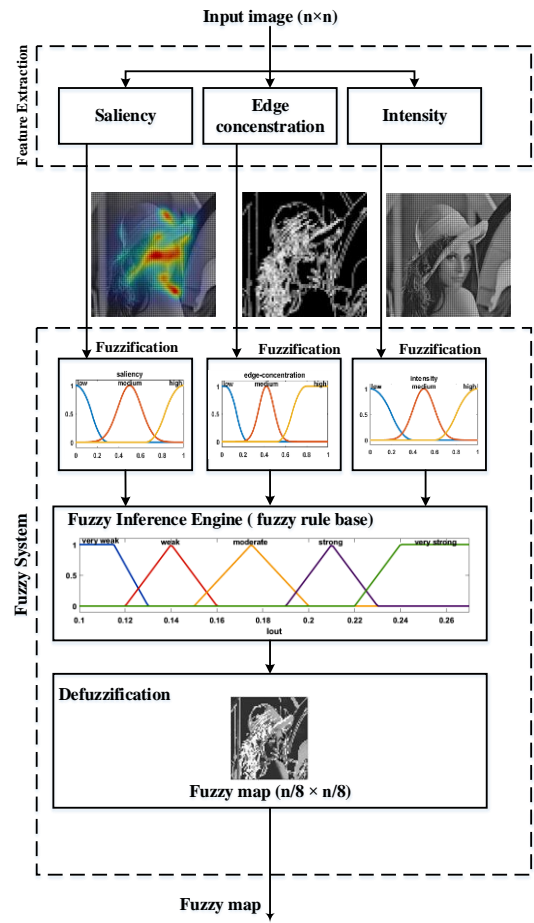

Fig. 1. Block diagram of the proposed feature extraction and fuzzy map formation.

### A. Features extraction

Here we consider features such as saliency, edge concentration, and intensity that could improve the embedding robustness. Therefore, we look for regions where the embedding has least perceptual effects. In the following, we explain those features.

*1) Saliency*

The main idea in saliency detection algorithms is to identify the most significant and visually informative parts of a scene. Salient parts are supposed to indicate human fixation locations of an image. To achieve imperceptibility in watermarking it is desirable to perform embedding in parts of an image that human eyes are less sensitive in their changes. Also, stronger data hiding in non-salient parts helps robustness. We use saliency method of [41]. This method has generated good results on our gray scale watermark dataset. Its output is a map with values in the range of [0 1] where salient parts have higher values. Figure 2(a) shows the original image. Fig. 2(b) shows the detected salient regions in the range of [0,1]. Also, in Fig. 2(c) a heat map, of the measured saliency values, shows the saliency map in which warmer colors present more salient regions.

*2) Intensity*

Based on Weber ratio concepts, changes in regions with high intensities could be unrecognizable as compared to dark areas. Hence, we try to have more powerful embedding in regions with high intensities than other regions. For this aim, first we divide image into $8 \times 8$ blocks. The average intensity of each block is calculated and normalized to [0 1].

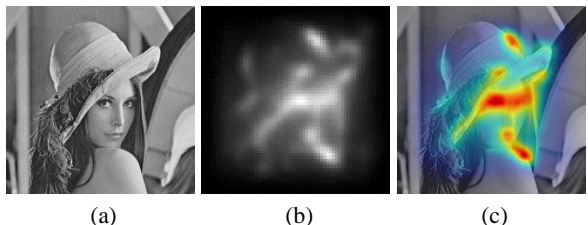

(a) (b) (c)

Fig. 2. Saliency map of [41]. (a) original image, (b) saliency map, (c) heat map of saliency.

*3) Edge-concentration*

Human visual system usually cannot recognize changes in parts of an image that are more irregular than other regions. Hence, such regions would be proper for powerful embedding. We find out regions with high concentration of edges as candidate areas for this aim. Therefore, we first use canny edge detector to almost find edges of image. Then regions with high edge density are selected. The process of edge-concentration is as follows:

1. Apply canny edge detector on the image.
2. Divide canny output into $8 \times 8$ blocks.
3. For every position in a block find variance of edge pixels around it in a $3 \times 3$ neighborhood.
4. Assign to each block the average of all variance values inside of that block.
5. Normalize this value between [0 1].

The obtained concentration-value of a block is then divided by this maximum value. Figure 3 shows the output of this phase.

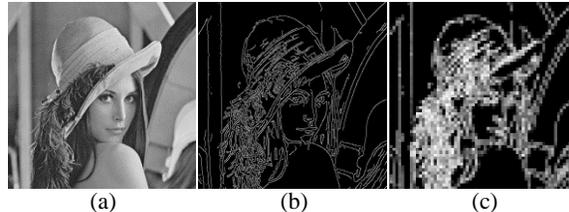

(a) (b) (c)

Fig. 3. Edge-concentration computation, (a) original image, (b) canny output map, (c) normalized edge-concentration map. This map is resized to $512 \times 512$.

As can be seen in Fig. 3, regions with more edges have higher concentration-values. Such blocks are more suitable for embedding purposes. Each input image has size $512 \times 512$. Hence, by dividing each image into $8 \times 8$ blocks and assigning the mentioned value to each block, we will have a $64 \times 64$ edge-concentration map. Fig. 3(c) is the resized version of this map to illustrate it with a better resolution.

*B. Fuzzy system*

Fuzzy inference system (FIS) is a method that assigns an output vector to each input vector, based on a set of predefined rules [42]. These rules are a list of if-then statements inspired from human experience. These rules can efficiently make decision like an expert [43]. In this method, we use Mamdani min-max fuzzy inference [44]. Saliency, intensity, and edge-concentration are fed to FIS as input attributes. Saliency, edge concentration and intensity are three fuzzy inputs. For these inputs, we determine three fuzzy term sets named low, high and medium with their special membership functions. The membership functions (MF), can be any function that maps each input space (referred to as the universe of discourse) to any value in the interval [0 1] and presents the degree of truth [42,45].

Most of the times, the choice of membership function depends on the problem. The membership function is determined heuristically, subjectively, or objectively [46]. In heuristic designs, selected membership functions may not reflect the actual data distribution [46]. Therefore, we do not use a heuristic approach. We try to design membership functions of the edge-concentration attribute objectively with a clustering method. This is done by generating probability density functions for each cluster to obtain the final membership functions. Moreover, membership functions for intensity have been designed subjectively with observation of experts and for saliency attribute have been designed with observations of volunteers

*1) Edge-concentration membership function*

For this input, designing of membership function is objectively done based on data distribution. Since we are introducing a new edge-concentration feature, it would be difficult to design a set of subjective membership functions. This is due to the lack of understanding of proper fuzzy intervals. For this aim, a clustering algorithm can be applied to estimate the actual data distribution. Finally, the resulting clusters can be used to produce the membership functions that will properly interpret the data [46].

In order to determine distribution of data on the edge-concentration feature, we first cluster the data using Fuzzy C-Means (FCM) clustering [47]. for clustering we considered 9 centers, experimentally 3 times more than our term-sets, seems to splitting the universe of discourse more densely. In Fig. 4(a), we fit a probability distribution to each of these clusters. Then we start to merge highly overlapped distributions that are next each other. This merging is done by considering the balance of data in each of the three newly merged distributions. Finally, we only need three probability density functions (PDF) to determine the mean ($\mu$) and variance ($\sigma$) of each of these fuzzy sets. This designing of membership function is more suitable for data partitioning. Figure 4(b) shows membership functions obtained from the 3 newly merged distributions. The average fuzzy-set is a Gaussian function with exact mean and variance of the related PDF. In other words, the related PDF has been scaled to become a normalized membership function with a maximum height of 1.

For the other two fuzzy-sets, i.e. high and low, the S-shaped and Z-shaped membership functions have been designed, respectively. In order to design these terminal functions, the peak-point and a bottom-point are needed. In our final membership functions, the peak point is the mean point of the related PDF and the bottom point is chosen according to the variance the distribution. We consider a Gaussian function of which one side stays at the peak level.

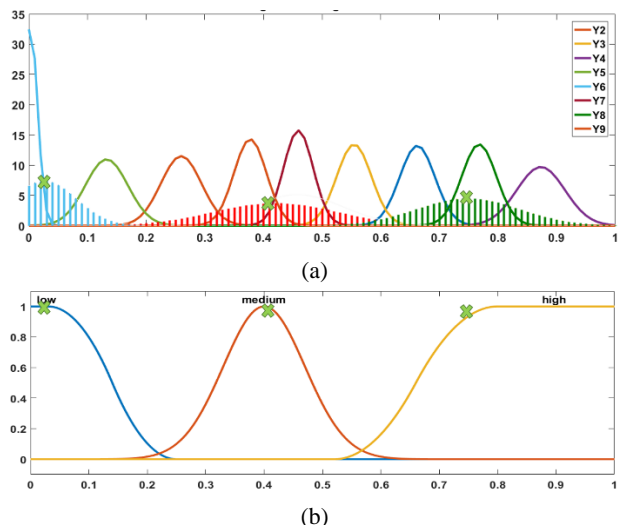

Fig. 4. Design of membership functions using FCM. (a) PDFs are fitted on 9 clusters, (b) final membership functions.

### 2) Saliency membership functions

For this feature, the distribution of data is determined using visual clustering in which three clusters were achieved by use of saliency heat map. Our three visual clusters are red to orange regions, yellow to blue, and gray areas. We consider these regions in standard watermark images according to heat map colors [41]. Then we sample the data in the corresponding box for each category. By using mean and variance of each category, we design membership functions. Figure 5 shows the process of building the saliency membership function.

### 3) Intensity membership functions

Intensity membership functions are designed totally subjectively based on SSIM results after embedding.

SSIM is dependent on the embedding strength factor. Embedding with high strength factor into DCT coefficients of a block will reduce SSIM. However, the SSIM value has smaller decay in blocks with high intensities. This complies with the human visual system. Hence, volunteers chose membership functions and decide on low, medium and high intensity values. The selection of membership function should cause least change in SSIM values.

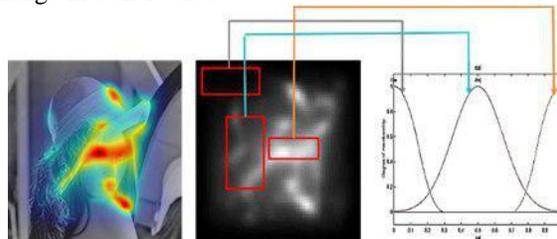

Fig. 5. Saliency membership function building.

### 4) Output membership functions

The fuzzy output map has five term-set consisting of very strong, strong, moderate, weak and very weak. All five fuzzy-set are designed subjectively by volunteers. Figure 6 shows the final output of the fuzzy membership function. This is built base on fuzzy rules that are explained later.

### 5) Fuzzy inference rules

For watermarking purpose, we design fuzzy rules based on saliency in different parts of an image. We consider high, medium, and low saliency areas. We have three attributes and each one has three values. Base on the combination of attributes and their values, we will have 27 rules that are subjectively generated by volunteers. Table I shows these FIS rules. As shown in this table, blocks with high saliency values have more detailed outputs due to human visual sensitivity to such areas. However, blocks with medium or low saliency have different rules due to lower visual sensitivity. Based on these rules, if a block has high saliency, data is embedded in it with low strength and the output fuzzy map would be very weak based on the intensity or edge-concentration features.

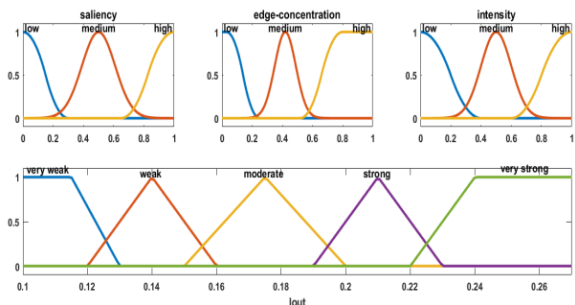

Fig. 6. Output membership function. Top row is input fuzzy membership function. Bottom row is output fuzzy membership function.

Table I. Summary of the proposed FIS fuzzy rules.

| Saliency | Edge-concentration | Intensity | Output |
|---|---|---|---|
| high | high | high | very strong |
| | | medium | strong |
| | | low | weak |
| | medium | high | strong |
| | | medium | weak |
| | | low | very weak |
| | low | high | moderate |
| | | medium | weak |
| | | low | very weak |
| medium | high | --- | very strong |
| | medium | not low | strong |
| | | low | weak |
| | low | high | strong |
| | | not high | weak |
| low | high | --- | very strong |
| | medium | not low | strong |
| | | low | moderate |
| | low | high | strong |
| | | not high | weak |

*6) Output map*

All watermark images in this study have the size of $512 \times 512$. Hence, there are $64 \times 64$ non-overlapped blocks of size $8 \times 8$ pixels. Defuzzified output of FIS is $64 \times 64$ image map. In the embedding phase, we use DCT of wavelet for data hiding. Details of the embedding process are explained in the next sub-section. Hence, each pixel of the fuzzy map determines the strength factor for its corresponding $8 \times 8$ block in DCT. Each pixel value of fuzzy map is in the range of $[0.1, 0.27]$. This map is considered as the final strength factor $\alpha$ that means, blocks with unsuitable conditions, which have low intensity, low edge-concentration, and high saliency, are assigned with the least strength factor of 0.1. This would be an initial strength factor and may be modified by a coefficient. Actually, FIS customizes the strength factor of each block using a fuzzy policy and by means of saliency, intensity, and crowdedness of pixels in a block, based on sensitivity of human visual system.

Figure 7 shows the output map that strength factor outputs are according to potential of each block for accepting strength factor in watermark embedding. The lighter pixels are the areas where stronger embedding will be done. For example, the fur in Lena's hat has high edge-concentration and medium saliency. This makes it suitable for very high strength factor (~ 0.27). Although Lena's mouth and eyes have high edge-concentration but human eyes first detect high saliency regions. Hence, embedding factor for Lena's mouth and eyes should be lower than the fur parts of the hat.

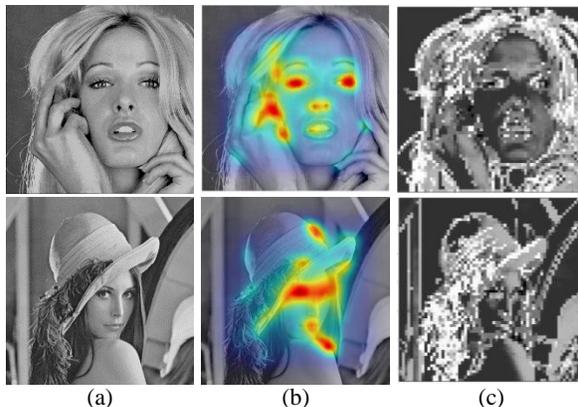

Fig. 7. Fuzzy strength factor map. (a) Original image, (b) saliency heat map, (c) fuzzy map.

### C. Watermarking scheme

Embedding in spatial domain is very vulnerable to attack and the embedded data is easily lost. Hence, spatial domain is not very suitable for data embedding and we use transform domain instead. DWT is one of the preferred domains that divides an image into different frequency levels. This division allows us to redundantly embed the data in different levels and hence improve the robustness of our method. For this aim we use a combination of DWT and DCT domains. We use HL, LH and HH sub-bands of a 2D wavelet transform in two levels. Figure 8 shows the selected sub-bands of wavelet domain with different colors.

Due to high sensitivity of image transparency to changes in LL coefficients, we do not use LL sub-band for embedding purpose. Redundant watermark into multiple parts help creating more robust embedding. Therefore, we redundantly embed in the seven sub-band levels, as shown in Fig. 8. In the following the embedding and extraction methods are detailed.

| | | LLHL | HLHL |
|---|---|---|---|
| LL | | | |
| | | LHHL | |
| LLLH | HLLH | LLHH | |
| LHLH | | | |

Fig. 8. Selected sub-bands for embedding purpose.

*1) Embedding algorithm*

The general steps of the embedding procedure are shown in Fig. 9.

For illustration purposes we consider that the images are $512 \times 512$ size and hence each designated sub-band, in Fig. 8, has the size of $128 \times 128$. But the proposed method is general and it could be applied to any size image.

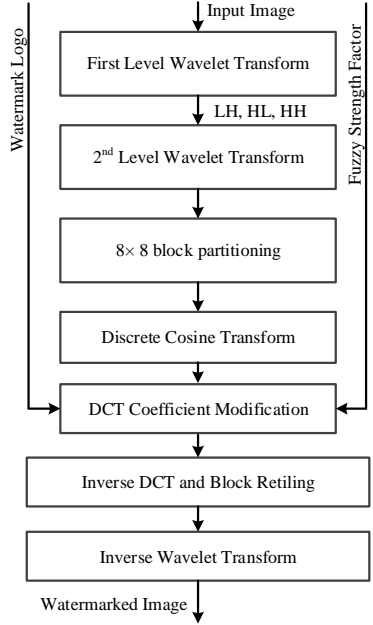

Fig. 9. Block diagram of the proposed embedding method

Each sub-band is partitioned into $8 \times 8$ blocks and DCT of each block is computed. Hence, we have 256 DCT blocks in each sub-band. For improving watermark robustness, we embed the watermark with redundancy. The watermark length is 128 bits while each sub-band has an embedding capacity of 256 bits. Therefore, the watermark is twice the embedded in each sub-band. Overall this 128-bit string is embedded 14 times in the whole image.

One bit of the watermark string is embedded into DCT coefficients of one $8 \times 8$ block. The embedding process to embed the *j*th copy of the *i*th watermark bit is performed in block $B_{ij}$ based on (1). Variable $W(B_{ij})$ shows the value of binary watermark bit that is to be embedded in block $B_{ij}$. Also $\alpha_{ij}$ is the adaptive strength factor for DCT block $(i,j)$. This strength factor is computed from the proposed fuzzy model. Two DCT coefficients of $D(u,v)$ and $D(x,y)$ are used for the embedding process, where $(u,v)$ and $(x,y)$ indicate the coordinates of the two coefficients in any $8 \times 8$ block. The relative value of these two coefficients shows whether there is a 1 or 0 embedded in this block. If $D(x,y) \geq D(u,v)$ we consider that a watermark bit of 1 already exists in this block, otherwise we assume a 0 is present in the block.

If, for example, a block by itself contains a 0 and we want to embed a watermark bit of 1 in this block, then the values of the coefficients are switched around to satisfy the condition. Figure 10 shows the pseudo code of the proposed embedding algorithm. In this pseudo code we are applying a strength factor of $\alpha_{ij}$. It means that if the values of the two coefficients of $D(u,v)$ and $D(x,y)$ are not different enough, the two coefficients are modified. The modified coefficients are guaranteed to be different enough based on the value of the strength factor $\alpha_{ij}$. This difference would help extraction of the watermark after signal processing attacks.

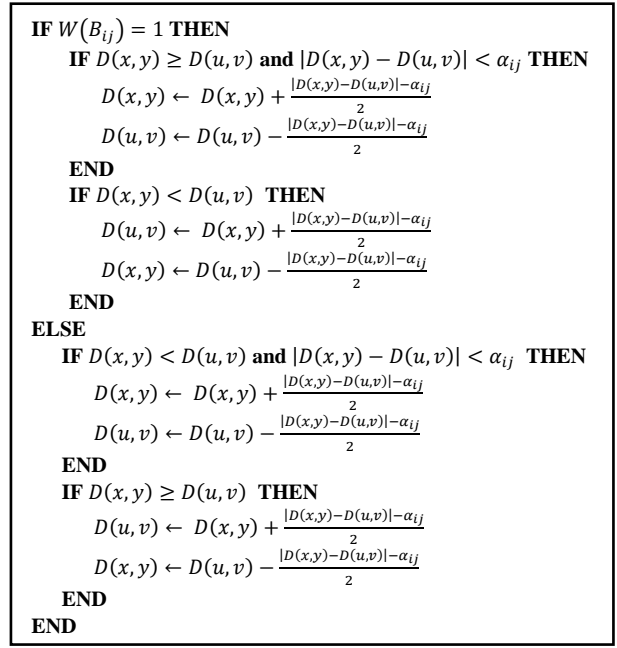

Fig. 10. Pseudo code of DCT coefficients modification for the proposed algorithm.

The mentioned steps of the pseudo code of Fig. 10 are repeated for all bits of the watermark string in each selected sub-band. After embedding a watermark bit in a 8×8 block of a sub-band, we perform inverse DCT of that block. At the end, when all designated sub-bands are embedded in, inverse DWT is performed to generate the watermarked image. This process of redundant embedding increases the robustness and improves the extracted logo's visual fidelity in the presence of attacks.

*2) Extraction scheme*

Extraction is done completely blind which could be considered as one of the advantages of the proposed scheme. Similar to embedding, wavelet transform is done in horizontal, vertical and diagonal sub-band of first level wavelet. After that DCT transformation are applied on

each block of seven selected sub-band. Extraction of the watermark string is performed based on equation (2):

$$W_x(B_{ij}) = \begin{cases} 1 & if \ D(x,y) \geq D(u,v) \\ 0 & if \ D(x,y) < D(u,v) \end{cases} \quad (2)$$

where $W_x(B_{ij})$ is the extracted bit from block $B_{ij}$. Ultimately, fourteen ($j = 1 \ to \ 14$) copies of the 128-bit ($i = 1 \ to \ 128$) string are extracted from the watermarked image. We use voting based on (3) to improve robustness of the algorithm

$$V_i = \sum_{j=1}^{14} W_x(B_{ij}) \quad (3)$$

where $V_i$ shows the result of vote counting for the $i$th bit of the 128-bit string, where $i = 1 \ to \ 128$. The voting is performed on all 14 versions of the extracted bit ($j = 1 \ to \ 14$). The vote count, $V_i$, has a value between [0 14]. The voting result is based on (4) by assigning 0 or 1 for each bit of the final extracted string:

$$\overline{W}_i = \begin{cases} 1 & if \ V_i > 6 \\ 0 & o.w. \end{cases} \quad (4)$$

where $\overline{W}_i$ is the final extracted watermark bit at position $i$ and $\overline{W}$ is the final extracted watermark string.

### III. EXPERIMENTAL RESULTS

In this section, several experiments have been done to evaluate the performance of proposed method. We use the common gray scale images that are used in [48]. All images are $512 \times 512$ and a randomly generated 128-bit string is used as the watermark. The Daubechies (db1) decomposition is performed for the wavelet transform. After taking DCT of each $8 \times 8$ block we select two of the coefficients. Experimentally we found out that coefficients at coordinates $(u,v) = (5,6)$ and $(x,y) = (6,5)$ have better robustness against most attacks. In addition, different coefficients of fuzzy map are used for each of the seven-mentioned wavelet sub-bands. Experimentally, coefficient values of 0.45, 0.45 and 0.1 have been chosen for the fuzzy map generation of HL, LH and HH levels respectively. The LLHH fuzzy coefficient is more sensitive to changes. Hence, we embed with a higher strength factor as compared to the other sub-bands.

The robustness and visual quality of our method is compared with some other existing watermarking methods, such as [28], [29], [37], [39] and [40] which have same conditions as ours. In order to have a fair comparison, the same message length is considered in our experiments.

#### A. Visual quality

Figure 11 shows results of applying our method to some standard images such as "Lena", "Couple", "Baboon" and "Lake". Peak signal-to-noise ratio (PSNR) is a simple and common metric for image quality. Due to ignoring human visual system, PSNR is not an accurate quality measure. Thus, we use both PSNR and Mean structural similarity index (MSSIM) for estimating perceptual quality of watermarked images. The proposed method keeps high perceptual quality of the watermark image. MSSIM value for all of the watermarked images shown in Fig. 11 is 1.

#### B. Robustness

To evaluate the robustness of our method, we calculate the normalized correlation ($NC$) between original and extracted watermark after different attacks. These attacks contain salt and pepper noise (S&P), JPEG compression, cropping, Gaussian filter, median filter and white noise. The $NC$ metric [49] was computed for each extracted string as follows:

$$NC = 1 - \frac{\sum_{i=1}^{M} |W(i) - \overline{W}(i)|}{M} \quad (5)$$

where $W$ is the original and $\overline{W}$ is the extracted watermark string and $M$ is the string length. In some of the mentioned methods the length of string is different than ours. Hence in our comparisons we consider this and if the string is smaller than ours, the same string length is used but with higher redundancy.

To measure robustness of our method, several attacks are applied to images and NC is computed for each string. Table II shows the NC results of our method against cropping attack. Different percentages of cropping (5%, 10%, 15%, 20%) are performed on mentioned image. Also, we considered two different cropping; corner of image and around the image. By increasing the percentage of cropping, our method has good robustness against attack in both kind of cropping.

Table II. NC results of our algorithm against cropping attacks.

| Image | | Barbara | | Baboon | | Couple | |
|---|---|---|---|---|---|---|---|
| | | C | A | C | A | C | A |
| NC | 5% | 1.000 | 1.000 | 1.000 | 1.000 | 1.000 | 1.000 |
| | 10% | 1.000 | 0.912 | 1.000 | 0.945 | 1.000 | 0.953 |
| | 15% | 1.000 | 0.912 | 1.000 | 0.945 | 1.000 | 0.938 |
| | 20% | 1.000 | 0.912 | 1.000 | 0.945 | 1.000 | 0.930 |

[1] Cropping center of the image.
[2] Cropping around the image.

Another important attack is Gaussian filter that we tested our method and evaluate its robustness. Three different window sizes $3 \times 3, 5 \times 5$ and $7 \times 7$ with different standard deviations ($\sigma = 0.5, 1, 2$) are used. The experimental results are indicated in Table III. High NC values shows the robustness of this method against Gaussian attacks. When $\sigma = 0.5$, our method has a good robustness regardless of the window size. By increasing the $\sigma$, the image is more smoothed and the frequency domain is more affected. Hence, the extracted watermark is destroyed and the NC value is reduced. But our method has acceptable robustness yet.

Table III. NC results of proposed method against Gaussian attack.

| image | $\sigma = 0.5$ | | | $\sigma = 1$ | | | $\sigma = 2$ | | |
|---|---|---|---|---|---|---|---|---|---|
| | 3×3 | 5×5 | 7×7 | 3×3 | 5×5 | 7×7 | 3×3 | 5×5 | 7×7 |
| Lena | 1.000 | 1.000 | 1.000 | 1.000 | 1.000 | 1.000 | 0.766 | 0.930 | 0.914 |
| Barbara | 1.000 | 1.000 | 1.000 | 0.969 | 0.992 | 0.992 | 0.820 | 0.875 | 0.898 |
| Baboon | 1.000 | 1.000 | 1.000 | 0.961 | 0.984 | 0.984 | 0.75 | 0.813 | 0.805 |
| Bridge | 1.000 | 1.000 | 1.000 | 0.961 | 0.938 | 0.953 | 0.695 | 0.773 | 0.711 |
| Couple | 1.000 | 1.000 | 1.000 | 0.984 | 0.992 | 0984 | 0.820 | 0.875 | 0.914 |

The robustness of our method against JPEG attacks is shown in Table IV. JPEG is one of the common attacks that our method has a good performance against it. Three standard images, such as Barbara, Lena and Boat, are used and bit error rate (BER) percentage values are computed. Low BER indicated the robustness of a method against these attacks.

Table IV. Comparison of BER (%) values for JPEG attack with different quality factors.

| image | Quality factor | | | | | | | |
|---|---|---|---|---|---|---|---|---|
| | 10 | 20 | 30 | 40 | 60 | 75 | 85 | 95 |
| Lena | 0.37 | 0.28 | 0.15 | 0.03 | 0.00 | 0.00 | 0.00 | 0.00 |
| Barbara | 0.29 | 0.19 | 0.08 | 0.01 | 0.00 | 0.00 | 0.00 | 0.00 |
| Baboon | 0.36 | 0.17 | 0.03 | 0.00 | 0.00 | 0.00 | 0.00 | 0.00 |
| Bridge | 0.22 | 0.07 | 0.03 | 0.02 | 0.00 | 0.00 | 0.00 | 0.00 |
| Pepper | 0.42 | 0.25 | 0.03 | 0.01 | 0.00 | 0.00 | 0.00 | 0.00 |

*C. Comparison with state of the art methods*

In order to evaluate the performance of the proposed method, we compare our method with recent algorithms of [28], [29], [37], [39] and [40]. These methods embed in DWT and CT respectively. All of these methods have used the same watermark length as we do. In each paper, some images with special attacks are considered. Hence, we compare our method with them based on their reported results. Bit error rates (BER) percentages are shown in Tables V to VIII. Lena, Barbara, Boat, Baboon, Couple, Bridge, Goldhill, Pepper, Pirate and Plane are some standard images that are used in this comparison. Different attacks, such as salt & pepper (1%, 3%, 5%), cropping (5%, 10%), Median filter (3 × 3), Gaussian filter (3 × 3), white noise, are considered here. These results clearly reveal that our method is highly robust against different mentioned attacks. Adaptive fuzzy strength factor helps us to have good robustness while it keeps a reasonable the degree of imperceptibility.

Tables V, VI, VII and VIII show comparisons between our method and methods of [28], [29], [37], [39] and [40]. In all comparisons, we try to produce images that have similar PSNR values similar to the mentioned references.

For attacks like salt & pepper, Gaussian noise, white noise, we repeat the experiments for 10 times and then the average of all BER values is indicated.

Table V shows the BER (%) values of method [37] and ours against salt & pepper attack. Different percentages of salt and pepper noise (1%, 3%, 5%) are examined. As we can see, in all cases our method performs better than the method of [37].

Table V. BER (%) values against Salt& Pepper attack for propose method, [37].

| Salt & pepper | 1% | | 3% | | 5% | |
|---|---|---|---|---|---|---|
| | ours | [37] | ours | [37] | ours | [37] |
| Barbara | 0.04 | 0.00 | **0.32** | 0.35 | **1.29** | 1.48 |
| Baboon | **0.00** | 0.63 | **0.38** | 0.63 | **0.71** | 2.89 |
| Couple | **0.15** | 0.59 | **0.62** | 3.05 | **5.28** | 8.71 |
| Bridge | **0.00** | 0.43 | **0.46** | 2.50 | **1.38** | 8.32 |

Based on Table VI, our method produces better results, in most cases, when exposed to the cropping attack. Cropping 5% and 10% are considered here.

Table VI. BER values against cropping attack for propose method, [40] and [29].

| cropping | 5% | | | 10% | | |
|---|---|---|---|---|---|---|
| | ours | [40] | [29] | ours | [40] | [29] |
| Lena | 1.53 | **0.78** | 1.83 | **4.69** | 5.85 | 5.92 |
| Goldhill | **0.00** | 6.64 | 2.84 | **4.69** | 13.67 | 6.53 |
| Bridge | **0.78** | 8.98 | 4.63 | **5.46** | 18.36 | 13.42 |
| Pepper | **0.00** | 3.51 | 6.58 | **4.69** | 6.25 | 10.48 |

In Table VII, results of [39] [40] are shown against Median and Gaussian filter attacks. Our method does not perform well under the median filter attack.

Table VII. BER values against median filter and Gaussian attacks for propose method, [39] [40].

| | Median filter 3 × 3 | | | Gaussian filter 3 × 3 | | |
|---|---|---|---|---|---|---|
| | ours | [40] | [39] | ours | [40] | [39] |
| Boat | 14.8 | 7.89 | **7.02** | 0.00 | 4.69 | 6.05 |
| Plane | 17.1 | 7.82 | **6.57** | 0.00 | 3.51 | 4.61 |
| Bridge | 18.7 | 11.72 | **6.93** | 0.00 | 5.86 | 4.59 |
| Pirate | 12.5 | 9.38 | **7.38** | 0.00 | 5.86 | 3.95 |

In Table VIII, we compare our method against different white noise attacks. Our method shows better results. The results are for the Lena image.

Table VIII. BER values against white noise (AWGN) attack in Lena for propose method, [28], [29] and [40].

| AWGN | $\sigma_n$ | | | | | |
|---|---|---|---|---|---|---|
| | 5 | 10 | 15 | 20 | 25 | 30 |
| ours | **0** | **0** | **0** | **0.05** | **2.34** | **4.49** |
| [28] | - | 1.85 | - | - | - | 26.39 |
| [40] | - | 5.01 | - | - | - | 20.11 |
| [29] | 3.34 | 8.23 | 13.84 | 18.45 | 23.58 | 27.52 |

As we can see, our method has better outputs against different attacks. In order to have a fair comparison, we consider the same watermark length in our experiments as the watermark length used in other works.

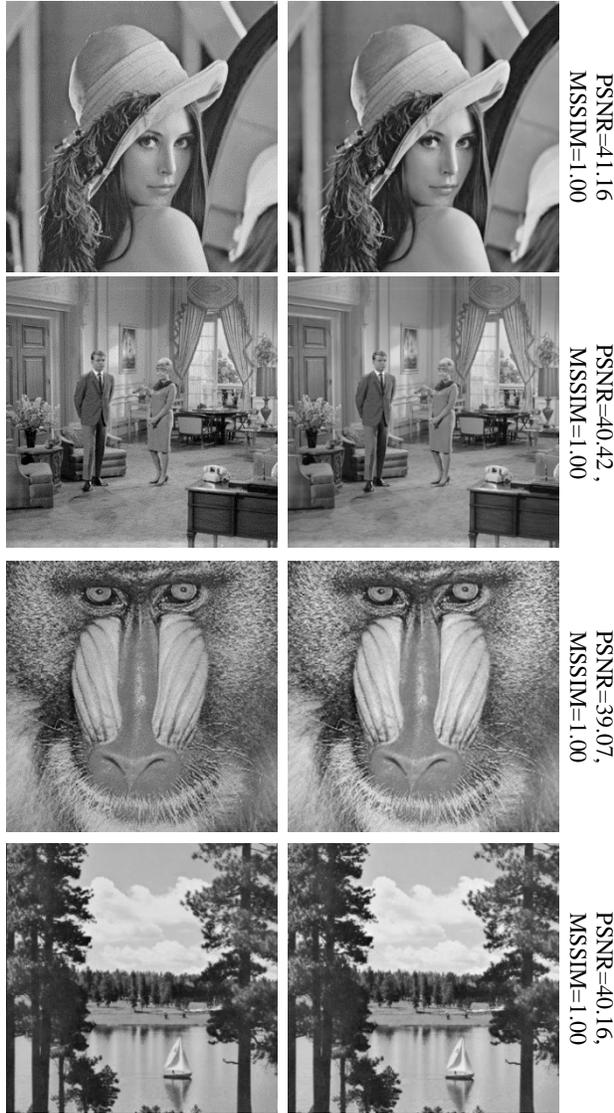

(a) original image      (b) watermarked image

Fig. 11. Comparison of the visual qualities between the original image and the watermarked image.

## IV. CONCLUSION

In this paper, we proposed a novel adaptive blind watermarking using fuzzy system. The proposed algorithm tries to use human psycho-visual characteristics to choose appropriate locations for embedding the watermark bits. At first a fuzzy inference system is fed with three attributes: saliency, intensity and edge-concentration of the original cover image. This FIS produces a map that used for adaptive embedding strength factor. The fuzzy system results sever embedding for regions that are not salient, contain edges, and have high intensities. After producing a fuzzy map, wavelet transform is applied on the original image and some of the second level sub-bands are used for embedding. Block level DCT of these sub-bands are computed and modification of some of the DCT coefficients would perfume the final embedding task. The combination of DWT and DCT resulted in high visual quality of the watermarked images. We tested our method against different attacks and experimental results show that this method has better performance in comparison with compared state-of-the-art methods.